\documentclass[useAMS,usenatbib]{mn2e}
\pdfoutput=1
\usepackage{epsfig}
\usepackage{graphicx}
\usepackage{times}
\usepackage{natbib}

\newif\ifAMStwofonts
\AMStwofontstrue

\newcommand{\be}{\begin{equation}}
\newcommand{\ee}{\end{equation}}
\newcommand{\ba}{\begin{eqnarray}}
\newcommand{\ea}{\end{eqnarray}}
\newcommand{\brr}{\begin{array}}
\newcommand{\err}{\end{array}}
\newcommand{\bc}{\begin{center}}
\newcommand{\ec}{\end{center}}

\newcommand{\hk}{\,h^{-1}{\rm kpc}}
\newcommand{\msun}{\,h^{-1}{\rm M}_\odot}
\newcommand{\sfr}{\,{\rm M}_\odot \, {\rm yr}^{-1}}
\newcommand{\hMpc}{\mbox{$h^{-1}{\rmn{Mpc}}~$}}

\newcommand{\mincir}{\raise
  -2.truept\hbox{\rlap{\hbox{$\sim$}}\raise5.truept \hbox{$<$}\ }}
\newcommand{\magcir}{\raise
  -2.truept\hbox{\rlap{\hbox{$\sim$}}\raise5.truept \hbox{$>$}\ }}
\newcommand{\siml}{\raise
  -2.truept\hbox{\rlap{\hbox{$\sim$}}\raise5.truept \hbox{$<$}\ }}
\newcommand{\simg}{\raise
  -2.truept\hbox{\rlap{\hbox{$\sim$}}\raise5.truept \hbox{$>$}\ }}

\title[Gas cooling in semi-analytic models and SPH simulations]
      {Gas cooling in semi-analytic models and SPH simulations: are results
        consistent?} 
\author[A. Saro, et al.]
       {A. Saro$^{1,2}$, G. De Lucia$^{3}$, S. Borgani$^{1,2,3}$ \& K. Dolag$^4$
         \\ 
         $^1$ Dipartimento di Astronomia dell'Universit\`a di Trieste, via
         Tiepolo 11, I-34131 Trieste, Italy (saro,borgani@oats.inaf.it)\\
         $^2$ INFN -- National Institute for Nuclear Physics, Trieste,
         Italy\\ 
         $^3$ INAF, Osservatorio Astronomico di Trieste, via Tiepolo 11,
         I-34131 Trieste, Italy (delucia@oats.inaf.it)\\
         $^4$ Max-Planck-Institut f\"ur Astrophysik, Karl-Schwarzschild Str. 1, 
         D-85748, Garching bei M\"unchen, Germany (kdolag@mpa-garching.mpg.de)\\
       }
\begin{document}

\date{Accepted ???. Received ???; in original form ???}

\maketitle

\label{firstpage}

\begin{abstract}
  We present a detailed comparison between the galaxy populations
  within a massive cluster, as predicted by hydrodynamical SPH
  simulations and by a semi-analytic model (SAM) of galaxy
  formation. Both models include gas cooling and a simple prescription
  of star formation, which consists in transforming instantaneously
  any cold gas available into stars, while neglecting any source of
  energy feedback. This simplified comparison is thus not meant to be
  compared with observational data, but is aimed at understanding the
  level of agreement, at the stripped-down level considered, between
  two techniques that are widely used to model galaxy formation in a
  cosmological framework and which present complementary advantages
  and disadvantages. We find that, in general, galaxy populations from
  SAMs and SPH have similar statistical properties, in agreement with
  previous studies. However, when comparing galaxies on an
  object-by-object basis, we find a number of interesting differences:
  {\em a)} the star formation histories of the brightest cluster
  galaxies (BCGs) from SAM and SPH models differ significantly, with
  the {\it SPH} BCG exhibiting a lower level of star formation
  activity at low redshift, and a more intense and shorter initial
  burst of star formation with respect to its {\it SAM} counterpart;
  {\em b)} while all stars associated with the BCG were formed in its
  progenitors in the semi-analytic model used here, this holds true
  only for half of the final BCG stellar mass in the SPH simulation,
  the remaining half being contributed by tidal stripping of stars
  from the diffuse stellar component associated with galaxies accreted
  on the cluster halo; {\em c)} SPH satellites can loose up to 90 per
  cent of their stellar mass at the time of accretion, due to tidal
  stripping, a process not included in the semi-analytic model used in
  this study; {\em d)} in the SPH simulation, significant cooling
  occurs on the most massive satellite galaxies and this lasts for up
  to 1 Gyr after accretion. This physical process is not included in
  the semi-analytic model used in our study, as well as in most of the
  models discussed in the recent literature. Our results identify
  specific directions of improvements for our methods to study galaxy
  formation in a hierarchical Universe.
\end{abstract}


\section{Introduction}
\label{sec:introduction}

In the current standard cosmological scenario, galaxies form when gas
that has been trapped in the potential wells of dark matter haloes and
has been shock heated to high temperatures, cools and condenses at the
centre of the haloes. The dynamical evolution of dark matter haloes is
governed by gravity alone and can be studied using either analytic
techniques (i.e. the Press--Schechter theory and its extensions;
\citealt{PressSchechter74,1991ApJ...379..440B,LA93.1}) or $N$-body
simulations \citep[e.g.][to mention just the most recent cosmological
simulation]{2009arXiv0903.3041B}. The evolution of the baryonic
component is, in contrast, less well understood and complicated by
gas-dynamical and radiative processes that can be treated only to some
extent using direct hydrodynamical simulations.

In fact, lacking a `complete theory' of star formation (as well as of
almost all the physical processes at play), we are currently not in
the position to model galaxy formation from first principles.

Different methods have been developed to model galaxy formation in a
cosmological context. Among these, semi-analytic models (SAMs) have developed
into a flexible and widely used technique to make detailed predictions of
galaxy properties. In these models, the physical processes driving galaxy
formation and evolution are approximated using physically and/or
observationally motivated analytic laws \citep[for a review,
  see][]{Baugh06}. Computational costs are limited so that this approach allows
an efficient investigation of the parameter space and of the influence of
different physical assumptions. Modern renditions of these models are coupled
to $N$-body simulations, thus removing uncertainties of the analytic approach
\citep{Benson_et_al_2005,Li_et_al_2007,Cole_et_al_2008} and providing
consistent dynamical information for model galaxies
\citep[e.g.][]{Kauffmann_etal_1999,Benson_etal_2000,Springel_etal_2001}. Another
approach consists in carrying out $N$-body + hydrodynamical simulations that
include both gas and dark matter
\citep[e.g.][]{KatzWeinberHernquist96,2001MNRAS.326..649P,2001ApJ...550L.129W}.
These simulations provide an explicit description of gas dynamics, but they are
still limited by relatively low mass and spatial resolution and by
computational costs that become highly prohibitive for simulations of galaxies
within large cosmological volumes. Recent work has analysed the properties of
galaxies in hydrodynamical simulations
\cite[e.g.][]{1996ApJ...472..460F,1999ApJ...521L..99P,2005ApJ...618...23N,2005ApJ...618..557N,2005MNRAS.361..983R,2006MNRAS.373..397S,2006MNRAS.373.1265O,
  Dave08}, but some difficulties are persistent. Albeit some recent progress
\citep[e.g.,][]{2008ASL.....1....7M}, it remains difficult for example to
produce realistic and rapidly rotating disks from cosmological initial
conditions \citep[e.g.][]{2009MNRAS.tmp..643S}.

In this paper, we will carry out a comparison between predictions of SAMs and
of SPH hydrodynamical simulations for the properties of galaxies inside and
around galaxy clusters. In order to minimize the uncertainties related to
different treatments of energetic feedback processes from supernovae and AGN,
we have considered only radiative gas cooling, and assumed that cooled gas is
instantaneously transformed into stars. It is important to stress that already
at this basic level, the different ways in which cooling is treated by SAMs and
hydrodynamical simulation codes may lead in principle to non-negligible
differences \citep[e.g.,][]{2008MNRAS.383..777V}.

Semi-analytic models assume that the haloes are spherically symmetric
and that the gas is efficiently shock heated to the `virial
temperature' of the halo. Hydrodynamical simulations do not require
any specific assumption on halo geometry, and solve explicitly the
hydrodynamic equations describing the evolution of the gas that is,
however, represented by a discrete number of fluid elements. Is is
difficult, given these limitations, to understand which of the two
techniques gives the correct answer.  While a comparison of the `full
semi-analytic model' with simulations that attempt to include all
physical processes at play would be of interest, a correct
interpretation of the results of such a comparison will necessarily
require a good understanding of any discrepancy induced by a different
treatment of gas cooling. In addition, it would be very difficult to
understand the origin of any similarity or difference between results
from non stripped-down versions of SAMs and simulations, as most of
the physical processes one would need to consider (e.g. feedback) are
necessarily included using quite different `recipes'.

Previous studies
\citep[e.g.][]{Benson_et_al_2001,Yoshida_et_al_2002,2003MNRAS.338..913H,Cattaneo_et_al_2007}
have compared numerical predictions from stripped-down versions of
semi-analytic models with those from hydrodynamical simulations to verify
whether these methods provide consistent predictions in the idealized case in
which only gas cooling is included.

In this study, we will present both a statistical comparison and an
object-by-object comparison between the two techniques. To this aim we use a
simulation of a massive cluster, and calculate the merger trees directly from
the simulation. While our results agree with those of previous studies in
showing that the two methods provide results that are statistically consistent,
we will highlight that significant differences arise when focusing on an
object-by-object comparison, particularly in the high-density environment of
galaxy clusters. As we will discuss below, some of the discrepancies
highlighted in this work have been already noted in previous studies, although
the general consensus is that the cooling model usually employed in SAMs is in
good agreement with hydrodynamical simulations that adopt the same physics.

The layout of the paper is as follows. In Sec.~\ref{simS2} we describe the
cluster simulation used in this study, while in Sec.~\ref{Sec:SAM2} we provide
a brief description of the adopted SAM. In Sec.~\ref{sec:Results1S2} we present
a statistical comparison between the cluster galaxy population predicted by the
two methods, while in Sec.~\ref{sec:Results2S2} we focus on a detailed
comparison of a sample of galaxies which have been matched in the different
runs. Finally, we discuss our results and give our conclusions in Sections
\ref{sec:disc} and \ref{concS2}.

%%%%%%%%%%%%%%%%%%%%%%%%%%%%%%%%%%%%%%%%%%%%%%%%%%%%%%%%%%%%%%%%%%%%%%%%%%%%%%%
\section{The simulation}
\label{simS2}

In this study, we use a re-simulation of a massive isolated galaxy
cluster with $M_{200}$ \footnote{We define $M_{200}$ as the mass
  contained within a radius encompassing a mean density equal to 200
  times the critical density of the Universe.}  $ \simeq 1.14 \times
10^{15} \msun$ and $r_{200}\simeq 1.7$ \hMpc, at $z=0$ (see also
cluster g51 from \citealt{2009MNRAS.399..497D}). The target cluster
was identified in a DM only simulation that followed the evolution of
$512^3$ DM particles (with a particle mass of $7\times
10^{10}\,h^{-1}\,{\rm M}_{\odot}$) in a comoving box of size
$479\,h^{-1}$Mpc on a side \citep{2001MNRAS.328..669Y}. The simulation
was carried out assuming a flat $\Lambda$CDM cosmology with
parameters: $\Omega_m = 0.3$, $h_{100} =0.7$, $\sigma_8 = 0.9$, and
$\Omega_b = 0.039$. All particles in the target cluster and its
immediate surroundings were then traced back to their Lagrangian
positions and re-simulated using the Zoomed Initial Condition (ZIC)
technique by \cite{TO97.2}, increasing the force and mass resolution
in the region of interest. In the high-resolution region, each DM
particle has a mass $m_{\rm DM}\simeq 1.13\times 10^9 \msun$, while
gas particles have mass $m_{\rm gas}\simeq 1.7 \times 10^8 \msun$ to
account for the cosmic baryon fraction.

The simulation was carried out using the TreePM--SPH code {\small GADGET-2}
\citep{Springel05}, and includes gravitational dynamics, gas cooling, and a
simple scheme of star formation, in which all gas particles colder than
$10^5$~K and denser than $4\times 10^{-27}$ g cm$^{-3}$ (corresponding to
$n_H=0.1$ cm$^{-3}$ for a gas of primordial composition) are immediately turned
into star particles. The simulation does not include a model for supernova
and active galactic nuclei (AGN) feedback. The Plummer--equivalent softening
length for the gravitational force was set to $\epsilon = 5 \hk$ in physical
units from $z=5$ to $z=0$, while at higher redshifts it was set to $\epsilon =
30 \hk$ in comoving units.  The smallest value assumed for the smoothing length
of the SPH kernel is half the gravitational softening.

Simulation data were stored in 93 outputs that are approximately
logarithmically spaced in time down to $z\sim 1$, and approximately
linearly spaced in time thereafter.  For each snapshot, we have run a
standard friends-of-friends (FOF) algorithm with a linking length of
0.2 in units of the mean particle separation. Each FOF group was then
decomposed into a set of disjoint substructures, identified by the
{\small SUBFIND} algorithm \citep{Springel_etal_2001} as locally over-dense
regions in the density field of the background main halo. For our
simulation, we have used a slight modification of the {\small SUBFIND}
algorithm (for details, see \citealt{2009MNRAS.399..497D}) which links
together all high-resolution particles (DM, gas, and star
particles). Only subhalos that retain at least 20 bound DM particles,
after a gravitational unbinding procedure, are considered `genuine
substructures'. We note that {\small SUBFIND} classifies all particles
inside a FOF group either as belonging to a bound substructure or as
being `unbound'. The self-bound part of the FOF group itself will then
also appear in the substructure list, and represents what we will
refer to below as the `main halo'. This particular subhalo contains
typically $\sim 90$ per cent of the mass of the FOF group
(e.g. \citealt{SpringelAl01}). The group catalogues were finally used to
construct merger histories of all gravitationally self-bound
structures using the software originally developed for the Millennium
Simulation
project\footnote{http://www.mpa-garching.mpg.de/galform/virgo/millennium/}. We
refer to \citet{2005Natur.435..629S} and to
\citet{2007MNRAS.375....2D} for a detailed description of the merger
tree construction algorithm. These merger trees represent the basic
input needed by the semi-analytic model used in our study (see Section
\ref{Sec:SAM2}). In the following, we will refer to all quantities
related to the simulation using the label {\it SPH}.

%%%%%%%%%%%%%%%%%%%%%%%%%%%%%%%%%%%%%%%%%%%%%%%%%%%%%%%%%%%%%%%%%%%%%%%%%%%%%%%
\section{The semi-analytic model}
\label{Sec:SAM2}

We use a stripped-down version of the semi-analytic model described by
\citet{2007MNRAS.375....2D}, which builds upon the methodology originally
introduced by \citet{Kauffmann_etal_1999}, \citet{Springel_etal_2001}, and
\citet*{2004MNRAS.349.1101D}.  In order to carry out a fair comparison with the
simulation described above, all physical processes in the semi-analytic model
have been switched off, with the exclusion of gas cooling and star formation.

Gas cooling is modelled following \cite{1991ApJ...379...52W}. The hot gas within
dark matter haloes is assumed to follow an isothermal profile:
\begin{displaymath}
\rho_{\rm g}(r) = \frac{M_{\rm hot}}{4\pi R_{200} r^2}
\end{displaymath}
Following \citet{2004MNRAS.349.1101D}, we define a cooling radius as
the radius at which the local cooling time is equal to the halo
dynamical time\footnote{We note that
  \protect\cite{1991ApJ...379...52W} defined the cooling radius
  equating the cooling time to the age of the Universe, which is about
  one order of magnitude larger than the halo dynamical time. As
  explained in \citet{2004MNRAS.349.1101D}, this particular choice was
  required by the significant enhancement of cooling rates in
  galaxy-size haloes when adopting metal dependent cooling
  functions.}. When the cooling radius lies within the virial radius,
the gas is assumed to cool quasi-statically and the cooling rate is
modelled by a simple inflow equation:
\begin{displaymath}
\frac{{\rm d}M_{\rm cool}}{{\rm d}t} = 4\pi\rho_{\rm g}(r_{\rm cool})r_{\rm
  cool}^2\frac{{\rm d}r_{\rm cool}}{{\rm d}t} 
\end{displaymath}
At early times and for low-mass haloes, the formal cooling radius is
larger than the virial radius. In these conditions, the hot gas is
never expected to be in hydrostatic equilibrium and the cooling rate
is essentially limited by the gas accretion rate. In this `rapid
cooling regime', we assume that all new diffuse gas that is added to
the halo, is accreted immediately onto the central object of the halo
under consideration. We note that different assumptions are made in
other published semi-analytic models (e.g. a cored gas profile
profile, and the inclusion of a free-fall time for gas to be
incorporated into the model galaxies). The model used in this paper,
however, has been shown to produce results that are in good agreement
with $N$--body $+$ hydrodynamical simulations that adopt the same
physics \citep{Yoshida_et_al_2002}. In addition, it provide results
that are in quite good agreement with other semi-analytic models which
adopts more realistic assumptions about the hot gas profile (De Lucia
et al., in preparation).

To simplify our comparison with simulation results, we have also
adopted a simplified star formation recipe: all gas that condenses
onto central objects via radiative cooling is immediately (rather than on a
disk dynamical timescale) turned into
stars. No other physical
process (e.g. supernova and AGN feedback, gas and/or metal recycling)
is considered.

We recall that the semi-analytic model adopted in this study includes
explicitly dark matter substructures: the haloes within which galaxies
form are still traced even when accreted onto larger systems. As
explained in \citet{Springel_etal_2001} and
\citet{2004MNRAS.349.1101D}, the adoption of this particular scheme
leads to the definition of three different `types' of galaxies. Each
FOF group hosts a `Type-0' galaxy, that is located at the position of
the most bound particle of the main halo, and is the only galaxy fed
by radiative cooling from the surrounding hot halo medium. All
galaxies attached to distinct dark matter substructures are referred
to as `Type-1'. These galaxies were previously central galaxy of a
halo that later merged onto the larger system in which they currently
reside. Positions of these galaxies are given by those of the most
bound particles of the subalos tracing the surviving cores of the
accreted haloes, while velocities are the mass weighted mean
velocities of all the selfbound particles. The hot reservoir that is
originally associated with the merging galaxy, is assumed to be
kinematically stripped at the time of accretion and is added to the
hot component of the remnant halo. Tidal truncation and stripping
rapidly reduce the mass of dark matter substructures below the
resolution limit of the simulation
\citep{2004MNRAS.348..333D,2004MNRAS.355..819G}. When this happens, we
estimate a residual surviving time for the satellite galaxies, using
the classical dynamical friction formula by \cite{1943ApJ....97..255C}
(see also \citealt{Saro08A}). The positions and velocities of these
galaxies are followed by tracing the most bound particles of the
subhalos at the last time they were identified, before being
disrupted.  Galaxies no longer associated with distinct dark matter
substructures are referred to as `Type-2' galaxies, and their stellar
mass is assumed not to be affected by the tidal stripping that reduces
the mass of their parent haloes.

We note that, by construction, the merger trees extracted from our simulation
will not include any Type-2 galaxy (only subhalos with more than 20 DM
particles are retained). In order to show how the semi-analytic results are
affected by the inclusion of a dynamical friction time-scale, we will analyse
in the following two different runs. We will use the label {\it SAM} to refer
to a run in which, whenever a subhalo falls below the resolution limit, its
galaxy is assumed to merge instantaneously with the galaxy sitting at the
centre of the main halo. This SAM run allows us to carry out a fair
object-by-object comparison between the simulation and the semi-analytic
results. We will also show results from a model that includes the dynamical
friction formulation described in \citet{2007MNRAS.375....2D} and
\citet{Saro08A}. These results will be indicated using the label {\it SAM2}. We
note that the SAM2 run should be compared with the SAM run, rather than with
the SPH simulation.

\begin{figure*}
\centerline{ 
\hbox{ 
\psfig{file=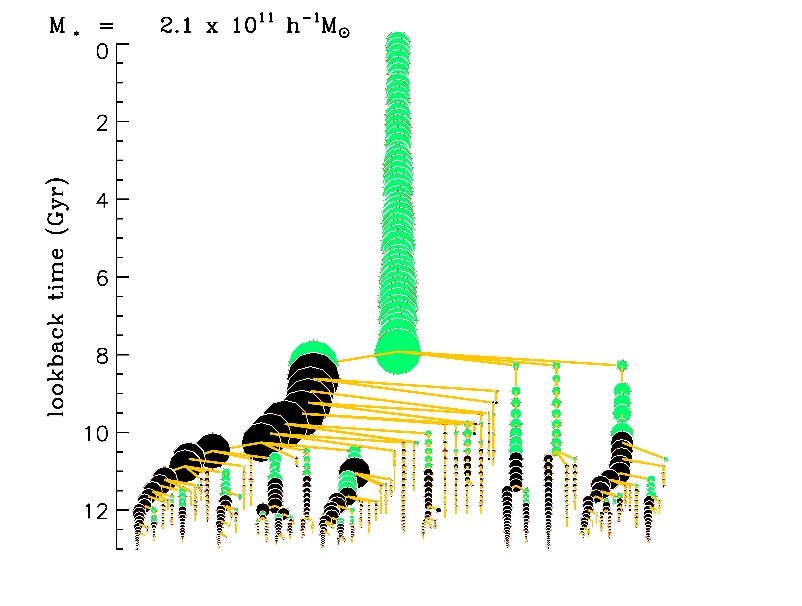,width=0.47\textwidth} 
\psfig{file=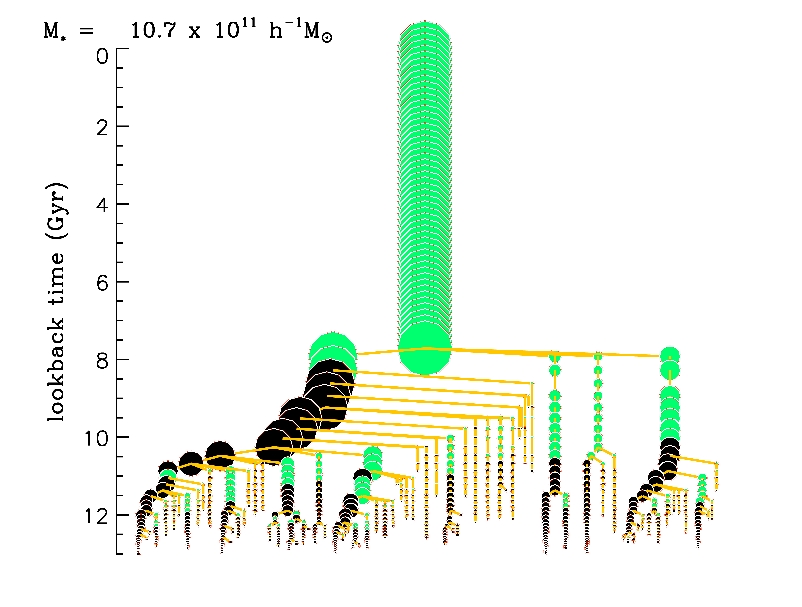,width=0.47\textwidth} 
 }} 
\centerline{ 
\hbox{ 
\psfig{file=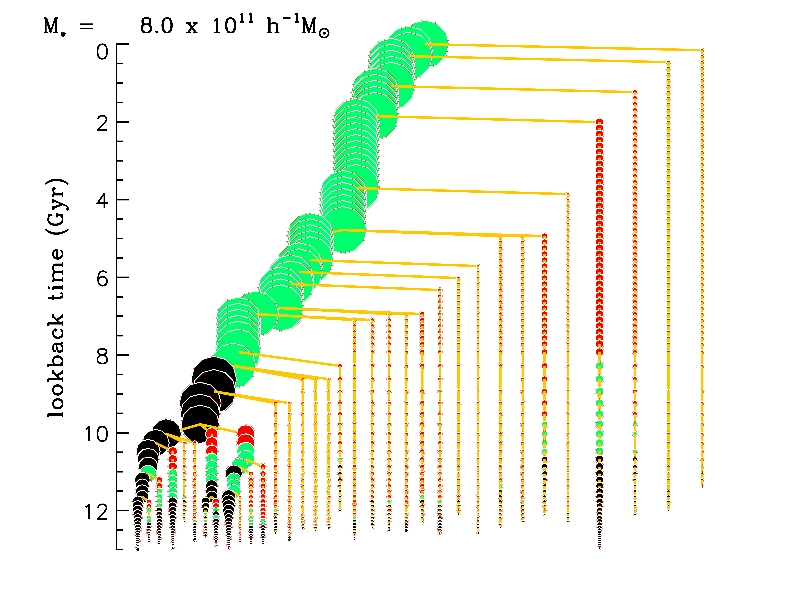,width=0.47\textwidth} 
\psfig{file=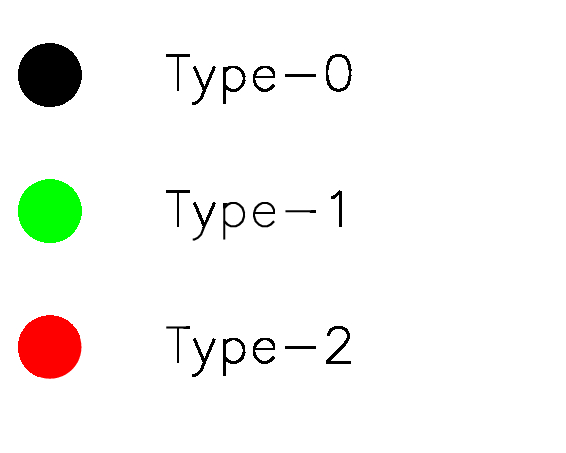,width=0.47\textwidth} 
 }} 
\caption{Galaxy merger tree of the sixth most massive satellite galaxy found in
  the simulation within $r_{200}$ at $z=0$. The upper left panel shows
  the galaxy merger tree extracted from the {\it SPH} simulation. The
  upper right panel and lower left panel show the corresponding merger
  trees from the {\it SAM} and {\it SAM2} runs (see text for
  details). The size of the symbols scales as the square root of the
  galaxy stellar mass, while different colours are used for different
  galaxy types (black for Type-0, green for Type-1, and red for Type-2
  galaxies - the latter exist only in the {\it SAM2} run). The final
  stellar mass of the galaxy in each run is given in the upper left
  corner of each panel.}
\label{fi:TreeS2}
\end{figure*}

To illustrate how results from the hydrodynamical simulation and from
the semi-analytic model used in this study compare to each other, we
show in Figure \ref{fi:TreeS2} the merger tree of the sixth most
massive satellite galaxy within $r_{200}$ at $z = 0$. The upper left
panel corresponds to the merger tree extracted from the {\it SPH}
simulation, while the corresponding trees from the {\it SAM} and {\it
  SAM2} runs are shown in the upper right and in the lower left panel,
respectively. In this figure, the size of the symbols scales as the
square root of the galaxy stellar mass, and different colours
correspond to different galaxy Types (black for Type-0, green for
Type-1, and red for Type-2 galaxies). We recall that by construction,
the {\it SPH} simulation and the {\it SAM} run do not contain Type-2
galaxies. This galaxy population is only present in the {\it SAM2}
run.

The final stellar masses of the {\it SAM} and {\it SAM2} galaxies are $M_*\simeq
10^{12} \msun$ and $M_*\simeq 8\times 10^{11} \msun$, respectively. The {\it SPH}
galaxy has a stellar mass of $M_*\simeq 2.1\times 10^{11} \msun$, lower than the
corresponding {\it SAM} and {\it SAM2} value by a factor five and four,
respectively. Figure \ref{fi:TreeS2} shows that the {\it SAM2} merger tree has
fewer branches than the corresponding {\it SPH} and {\it SAM} trees. This
happens because of the large number of surviving Type-2 galaxies which are
progenitors of the final galaxy in the {\it SPH} and {\it SAM} runs as we set
the merging times equal to zero once the parent substructure falls below the 20
DM particles limit. In the {\it SAM2} run, these galaxies are assigned a
residual merging time which is longer than the time interval between the
lookback time corresponding to their `appearance' and present, and are
therefore not included in the merger tree shown in the bottom panel of Figure
\ref{fi:TreeS2}.

The figure shows that in the {\it SAM} and {\it SAM2} runs, galaxies increase
their stellar mass mostly while they are Type-0 central galaxies and gas is
allowed to radiatively cool towards their centres to form stars. Satellite
Type-1 galaxies can increase their stellar mass only via merging. In the
semi-analytic model, their stellar mass never decreases as stellar stripping is
not modelled. In contrast, in the {\it SPH} simulation, satellite Type-1
galaxies continuously loose mass because of tidal interactions with the main
halo. The stellar material that is lost is deposited in a diffuse stellar
component that is associated with the central galaxy of the main halo by
{\small SUBFIND}. By the time satellite galaxies merge in {\it SPH} simulation,
their stellar masses are significantly reduced. We will come back to this issue
in Sections \ref{sec:Results1S2} and \ref{sec:Results2S2}.

\begin{figure*}
\centerline{ \hbox{ \psfig{file=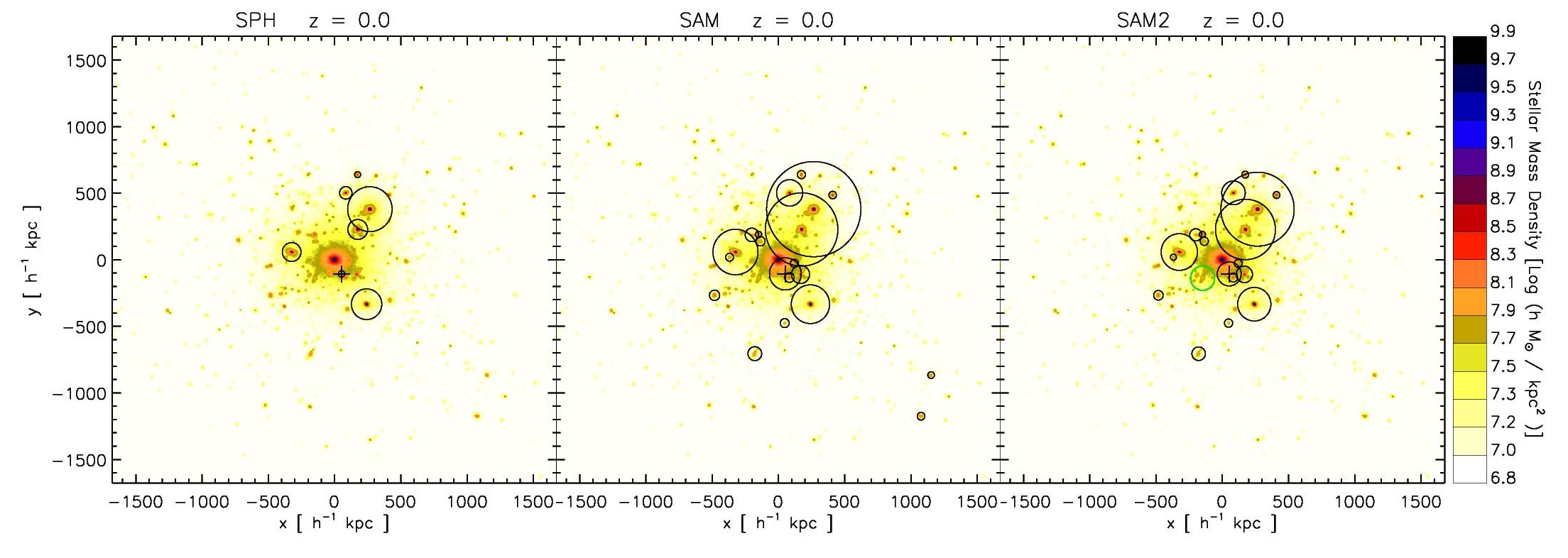,width=20cm}}}
 \caption{Stellar density maps of a box centred on the main halo of
   the cluster used in this study and of size $2 \,r_{200}$ on a
   side. The positions of all the satellite galaxies with
   stellar mass larger than $2 \times 10^{11} \msun$ are marked by
   circles, with larger radii corresponding to larger stellar
   masses. The left panel corresponds to the {\it SPH} run, the
   central panel is for the {\it SAM} run, and the right panel is for
   the {\it SAM2} run. Different colours are used for different galaxy
   types: black for Type-1 and green for Type-2 galaxies (only present
   in the {\it SAM2} run). Position of the sixth most massive
   satellite galaxy in the simulation shown in Fig. 1 is highlighted
   with a cross.}
  \label{fi:MapsS2}
\end{figure*}

Figure \ref{fi:MapsS2} shows the stellar mass density within a box of
$2\, \,r_{200}$ on a side, centred on the main halo of the
cluster. Superimposed on each map are the positions of all the
satellite galaxies with stellar mass larger than $2\times 10^{11}
\msun$. Each satellite galaxy is indicated by an open circle whose
radius scales with the galaxy stellar mass. Each panel corresponds to
a different run (left to {\it SPH}, central to {\it SAM}, and right to
{\it SAM2}). Only one Type-2 galaxy is found in this region in the
{\it SAM2} run, and is marked with a green circle. Position of the
sixth most massive galaxy in the simulation shown in Fig. 1 is marked
with a cross. The figure shows that the satellite galaxies in the {\it
  SPH} run are significantly less massive than those in the {\it SAM}
and {\it SAM2} run, because of the tidal stripping of stars discussed
above. As a consequence, semi-analytic runs contain a larger number of
massive galaxies within the virial radius.

%%%%%%%%%%%%%%%%%%%%%%%%%%%%%%%%%%%%%%%%%%%%%%%%%%%%%%%%%%%%%%%%%%%%%%%%%%%%%%%
\section{A statistical comparison}
\label{sec:Results1S2}

In this Section, we will carry out a statistical comparison between the
properties of galaxies from the {\it SPH} simulation and those from the {\it
  SAM} and {\it SAM2} runs. In particular, we will focus on the radial density
distribution and on the stellar mass function. In the next section, we will
then carry out a more detailed object-by-object comparison.

\begin{figure}
\centerline{ \hbox{ 
\psfig{file=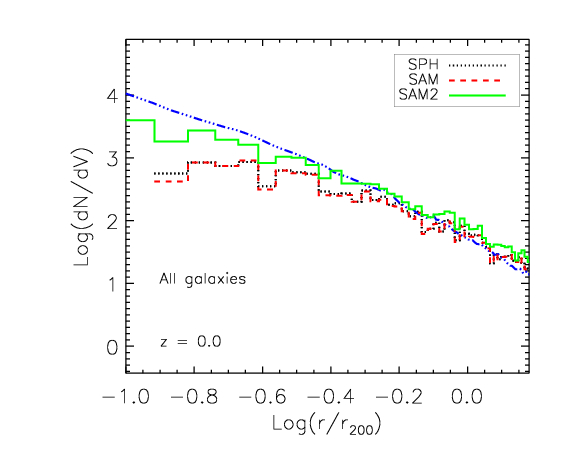,width=0.5\textwidth}
}} 

\caption{Radial density distribution of all galaxies identified at $z
  = 0$ within $1.5\, r_{200}$. The dotted black line shows the
  distribution of galaxies identified in the {\it SPH} simulation,
  while the red and green lines show the corresponding results from
  the {\it SAM} and {\it SAM2} runs. The dot-dashed blue line shows
  the total matter density profile, normalised to match the radial
  number density of galaxies in the {\it SAM2} run at $0.5\,
  r_{200}$.}

\label{fi:rad_profS2}
\end{figure}

Figure \ref{fi:rad_profS2} shows the radial density distribution of
all galaxies lying within $1.5 \,r_{200}$ at $z = 0$. The dotted black
line shows the distribution of galaxies identified in the {\it SPH}
simulation, while the dashed red and solid green lines corresponds to
the {\it SAM} and {\it SAM2} runs respectively. The dot-dashed blue
line shows the total mass density profile, normalised to match the
radial number density of galaxies in the {\it SAM2} run at $0.5 \,
r_{200}$. As shown by \citet{2004MNRAS.352L...1G}, a run that includes
Type-2 galaxies traces better the mass density profile, while the
corresponding distributions from the {\it SPH} and {\it SAM} runs are
anti-biased with respect to the matter distribution in the inner
regions, as the subhalo profiles \citep{2004MNRAS.348..333D,
  2004MNRAS.355..819G}. Interestingly, the contribution from the
Type-2 population in the {\it SAM2} run remains significant also at
large radii, and even beyond the cluster virial radius.

Figure \ref{fi:rad_profS2} indicates a small excess of {\it SPH} satellites
with respect to the {\it SAM} results, in the inner central bin. We have
verified that this small excess is due to galaxies which do not have any
counter-part in the {\it SAM} run because of a failure of the merger tree
reconstruction algorithm.  In order to maximize the algorithm performance for
haloes that are close to the 20 particles limit, whenever a descendant of a
halo under consideration is not found in the next simulation snapshot, the
algorithm also searches for a descendant in the subsequent snapshot. Close to
the resolution limit, however, there are a few haloes that disappear for more
than two contiguous snapshots. These are identified at a later time as satellite
galaxies with a non-zero stellar mass in the simulation. In the semi-analytic
model, gas is allowed to cool only on central galaxies of `main haloes'. Since
these structures appear as `subhalos', no new galaxy is formed within
them. There will be, however, a type 2 galaxy associated with the (erroneously)
lost Type-1 galaxy. We have verified that this `failure' is actually occurring
for all Type-2 galaxies in the {\it SAM2} run whose associated merging times
are large enough to let them survive as distinct galaxies down to $z=0$. The
positions assigned to the Type-2 galaxies coincide with those of the Type-1
{\it SPH} galaxies which are erroneously identified as new.

\begin{figure*}
  \centerline{ \hbox{ \psfig{file=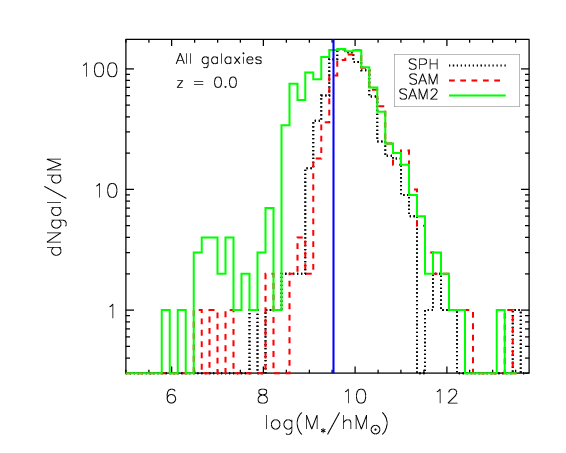,width=0.5\textwidth} \psfig{file=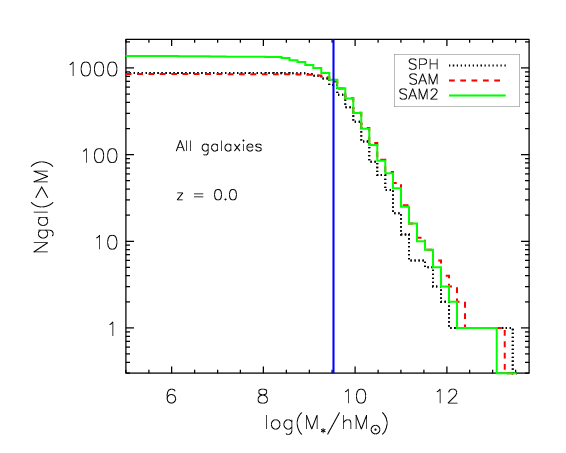,width=0.5\textwidth}
    }} 
\caption{Differential (left panel) and cumulative (right panel) stellar mass
  functions for all galaxies $z=0$ within $1.5 \,r_{200}$ of the simulated
  cluster used in our study, as predicted by the {\it SPH} simulation (black
  dotted lines), by the {\it SAM} run (red dashed lines) and by the {\it SAM2}
  run (green solid lines). The solid vertical blue lines mark the resolution
  limit of the simulation.}
\label{fi:MFSS2}
\end{figure*}

Figure \ref{fi:MFSS2} shows the stellar mass function of all galaxies
identified within $1.5 \, r_{200}$ at $z = 0$. The differential and
cumulative mass functions are shown in left and right panel,
respectively.  Dotted black lines represent the stellar mass function
of galaxies identified in the {\it SPH} run, while red and green lines
correspond to the {\it SAM} and {\it SAM2} runs, respectively. The
solid vertical blue lines in the two panels mark the mass
corresponding to the cosmic baryonic fraction multiplied by the mass
of 20 DM particles, and therefore roughly corresponds to the
resolution limit of the simulation.

The figure shows that the stellar mass of the {\it SAM} BCG is larger than the
corresponding value in the {\it SAM2} run, but smaller than the corresponding
value in the {\it SPH} run. We recall that the latter value includes the
intra-cluster light. The difference between the {\it SAM} and {\it SAM2} BCGs
is mainly due to the lower number of progenitors of the {\it SAM2} BCG (see
Figure \ref{fi:TreeS2}), which translates in a much larger (by about 50 per
cent) number of galaxies at $z=0$ with respect to the {\it SAM} (and {\it SPH})
run, as clearly shown by the cumulative mass functions.  As explained above,
there are more massive galaxies in the {\it SAM} run than in the {\it SPH}
simulation. As we will see in the following, the differences between the {\it
  SPH} and the semi-analytic runs can be explained as a combination of tidal
stripping of stars from the {\it SPH} satellite galaxies, and to different
predicted star formation histories for the BCGs.

%%%%%%%%%%%%%%%%%%%%%%%%%%%%%%%%%%%%%%%%%%%%%%%%%%%%%%%%%%%%%%%%%%%%%%%%%%%%%%%
\section{An object-by-object comparison}
\label{sec:Results2S2}

The dynamical evolution of central galaxies and of satellite Type-1 galaxies in
the semi-analytic model is based on merger trees that are equivalent to those
extracted from the {\it SPH} simulation. This allows us to identify a sample of
galaxies which includes all those at $z=0$ that have the same positions in all
three runs. This sample includes all central Type-0 galaxies and all satellites
Type-1 galaxies, with the exception of those Type-1 satellites in the {\it SPH}
run which do not have a counterpart in the {\it SAM} and {\it SAM2} runs as
described in Section \ref{sec:Results1S2}. 

%%%%%%%%%%%%%%%%%%%%%%%%%%%%%%%%%%%%%%%%%%%%%%%%%%%%%%%%%%%%%%%%%%%%%%%%%%%%%%%
\subsection{General behaviour}

\begin{figure*}
  \centerline{ \hbox{
      \psfig{file=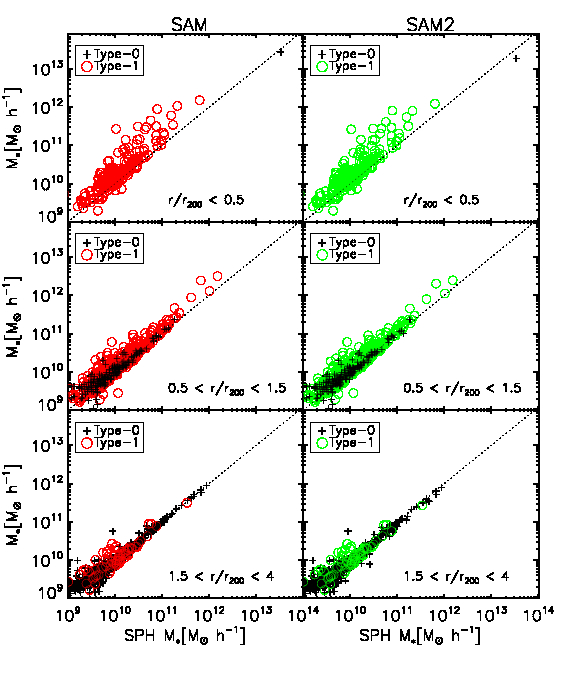,width=\textwidth}}} 
\caption{Stellar mass of model galaxies from the {\it SAM} (left
  panels) and {\it SAM2} (right panels) runs as a function of the
  corresponding stellar mass as measured from the {\it SPH}
  simulation. Panels from top to bottom show bins of increasing
  cluster-centric distance. Central Type-0 galaxies are shown as black
  crosses, while red and green circles represent satellite Type-1
  galaxies in the {\it SAM} and {\it SAM2} runs, respectively. The
  dotted line in each panel represents the one-to-one relation and is
  shown to guide the eye.}
\label{fi:Mass_comparison_radiusS2}
\end{figure*}

Figure \ref{fi:Mass_comparison_radiusS2} shows the stellar mass of
matched galaxies at $z=0$ as predicted by the {\it SPH} simulation
(x-axis), by the {\it SAM} run (y-axis, left panels), and by the {\it
  SAM2} run (y-axis, right panels). Type-0 central galaxies are shown
as black crosses, while Type-1 satellite galaxies are shown as red
circles for the {\it SAM} run and as green circles for the {\it SAM2}
run. The dotted line in each panel represents the one-to-one
relation. Different panels correspond to different radial bins, with
cluster-centric distance increasing from top to bottom panels.

In the innermost region of the cluster (upper panels), corresponding to
galaxies at distances smaller than $0.5 \, r_{200}$, there is only one
Type-0 central galaxy, which is the BCG itself. On average, satellite galaxies
in the {\it SAM} and {\it SAM2} runs tend to be more massive than in the {\it
  SPH} simulation, even by more than one order of magnitude. The scatter above
the one-to-one relation is also larger in the innermost radial bin. For central
Type-0 galaxies, the agreement between the simulation and the semi-analytic
runs is better. We note, however, that the {\it SPH} BCG has a stellar mass
that is $\sim 20$ and $\sim 65$ per cent larger than the corresponding values
in the {\it SAM} and {\it SAM2} runs, respectively. In the next section, we
will show that these differences are mainly due to tidal stripping of stars
from satellite galaxies in the {\it SPH} simulation. At larger distances from
the cluster centre, the importance of tidal stripping decreases and the
agreement between the {\it SPH} and the semi-analytic runs improves.

%%%%%%%%%%%%%%%%%%%%%%%%%%%%%%%%%%%%%%%%%%%%%%%%%%%%%%%%%%%%%%%%%%%%%%%%%%%%%%%
\subsection{Evolution of the Brightest Cluster Galaxies}
\label{sec:bcgs}

As discussed above, the {\it SPH} simulation predicts for the BCG a
stellar mass that is larger than the corresponding values found in the
{\it SAM} and {\it SAM2} runs.  In this section, we discuss in more
detail the origin of this differences. As explained above, the BCG
mass in the SPH simulation includes the diffuse stellar component. The
semi-analytic model used here does not account for tidal stripping of
stars from satellite galaxies and does not model explicitly the
formation of the ICL. However, \citet{2007MNRAS.377....2M} have shown
that most of the diffuse stellar component does not come from tidal
stripping but originates from mergers associated with the formation of
the BCG. Thus the {\it SAM} BCG will also include the ICL, with the
exclusion of the (small) contribution of stellar material that was
tidally stripped from satellite galaxies.

In Figure \ref{fi:GAL0S2}, we show the mass accretion and star
formation history of the BCG as predicted by the {\it SPH} simulation
(black lines), by the {\it SAM} run (red lines) and by the {\it SAM2}
run (green lines). The upper left panel shows the number of
progenitors of the BCG as a function of lookback time for the three
runs. As explained in Section \ref{Sec:SAM2}, the number of
progenitors in the {\it SPH} simulation and in the {\it SAM} run is
larger than in the {\it SAM2} run. This difference is due to the
Type-2 galaxies, which maintain their identity down to $z=0$ in the
{\it SAM2} run. By construction, this population of galaxies does not
exist in both the direct simulation and in the {\it SAM} galaxy
catalogue, which are based on equivalent merger trees. In these runs,
whenever a substructure `merge' (i.e. falls below the resolution limit
of the simulation), the associated galaxy is assumed to merge
instantaneously with the central galaxy of its halo. Thus, the number
of progenitors is higher in these two runs. The merging rate (the
derivative of the curve shown in the top left panel of Figure
\ref{fi:GAL0S2}) predicted by the {\it SPH} and by the {\it SAM} runs
looks quite different from the one predicted by the {\it SAM2} run. In
particular, the merging rate is almost linear along the whole galaxy
history in the {\it SAM2} run, while in the {\it SPH} simulation and
in the {\it SAM} run it is characterised by a phase of high merging
activity before lookback time $\sim 7$ Gyrs, followed by a shallow
decline which lasts down to a lookback time $\sim 2$ Gyrs, and then
another phase of high merging activity in the past 2 Gyrs.

The upper right panel of Figure \ref{fi:GAL0S2} shows the star
formation history of the BCG (obtained by summing up the star
formation histories of all its progenitors). The star formation
histories predicted by the three models differ significantly. The {\it
  SPH} BCG exhibits an intense initial burst of star formation at
lookback time $\sim 12$ Gyrs, followed by a sharp decline. After
lookback time $\sim 8$ Gyrs, the star formation rate associated with
the BCG is roughly constant at a value of $\sim 450 \msun$
yr$^{-1}$. The corresponding star formation
histories from the semi-analytic model look very different. As for the
{\it SPH} BCG, the {\it SAM} run exhibits an initial burst of star
formation at early times. This burst is, however, less intense and
broader than the one predicted by the {\it SPH} simulation. The
different intensity of the initial burst of star formation might be
due to inaccurate/simplified treatment of the `rapid cooling
regime'. In the semi-analytic model used in this study, all hot gas
associated with the main halo is assumed to condense instantaneously
onto the central galaxy when the formal cooling radius is larger than
$r_{200}$ (see Section~\ref{Sec:SAM2}). This simple implementation
does not seem to describe accurately the deposition of cold gas along
filaments that penetrate the virial radius without being heated to the
virial temperature (\citealt{Dekel08} and references therein). 

In the last $\sim 10$ Gyrs, the star formation history of the {\it SAM} BCG is
roughly constant with an average star formation rate of $ \sim 2000 \msun$
yr$^{-1}$. For the {\it SAM2} BCG, the star formation history appears roughly
constant along the whole history of the galaxy with an average value of $\sim
2000 \msun$ yr$^{-1}$. It is close to the values predicted in the {\it SAM} run
in the last $\sim 10$ Gyrs, but is significantly lower than the corresponding
values predicted in the {\it SAM} run and {\it SPH} simulation at early
times. This difference can be ascribed to the lower number of progenitors of
the {\it SAM2} BCG, as shown in the upper left panel. Indeed, the Type-2 galaxy
population, which is present in the {\it SAM2} run, has formed its stellar mass
at very early times.

As explained above, our comparison is based on a stripped-down version
of the semi-analytic model and on a simulation in which star formation
is treated in a very similar way. Hence, any difference in the
predicted star formation histories reflects a difference in the amount
of gas that cools in the model and in the simulation. The `bursty'
behaviour of the star formation histories predicted in the {\it SAM}
and {\it SAM2} runs is therefore due to differences in the predicted
cooling rates, and not to bursts induced by mergers. As explained in
Section \ref{Sec:SAM2}, the amount of gas that cools ($M_{cool}$) in
the semi-analytic model is proportional to the mass of gas enclosed
within the cooling radius ($r_{cool}$). The hot gas is assumed to
follow an isothermal profile, so the density of gas at a radius $r$
scales as $r^{-2}$. Therefore, the total mass of gas within a certain
radius $r$ is $M_{gas}(<r) \propto r$.  In absence of any feedback,
gas particles in the SPH simulation assume a radial profile that is
not well described by an isothermal distribution (see
\citealt{2006MNRAS.367.1641B}). Figure \ref{fi:prof_dens} shows the
DM, star and gas mass enclosed within a given radius as a function of
cluster-centric distance. The innermost regions of the cluster
($r\mincir 0.01 \,r_{200}$) are dominated by the stellar component
(dashed red line), while the outer regions are dominated by DM (black
solid line). Satellite galaxies and the diffuse intra-cluster light
associated with the BCG make the stellar component contributing more
than the gas component at $r \sim 0.1 \, r_{200}$.  At larger radii,
the gas component represents the dominant baryonic component. The
dotted blue line in Figure \ref{fi:prof_dens} represents the resulting
gas profile assumed to compute the gas cooling rate in the SAM
runs. At the virial radius $r_{200}$, the gas mass measured from the
{\it SPH} simulation agrees very well with the gas mass predicted by
the {\it SAM} run.  For radii smaller than $\sim 0.2 \,r_{200}$, the
gas distribution assumed in the semi-analytic model differs
significantly from the actual gas distribution in the
simulation. Here, cooling at high redshift has already removed low
entropy gas, leaving only a small amount of gas able to cool at lower
redshift.  At $z = 0$, our semi-analytic code computes a cooling
radius for the main cluster of $\sim 40 \hk$. Figure
\ref{fi:prof_dens} shows that, within this radius (marked by a
vertical dotted line), the total gas mass assumed by the semi-analytic
model is about a factor four larger than the gas mass present in the
{\it SPH} simulation. The predicted cooling rate at $z = 0$, and
therefore the predicted star formation rate in the {\it SAM} run, will
be about a factor four larger than in the {\it SPH} run, as shown in
the upper right panel of Fig. \ref{fi:GAL0S2}. We note that not all
semi-analytic models (e.g. \citealt{ColeAl00} and derivative papers)
use this approximation. The different behaviour between semi-analytic
models and hydrodynamical simulations, however, is exacerbated in runs
without feedback, like those considered in our analyses. When a model
for efficient supernovae and/or AGN feedback is included, a larger
amount of gas is allowed to remain in the hot phase in the central
regions, thus possibly alleviating the disagreement shown in Figure
\ref{fi:prof_dens}.

The lower left panel of Figure \ref{fi:GAL0S2} shows the total mass in all BCG
progenitors, in the three runs considered in our study. In the {\it SPH}
simulation, there is a first phase (at lookback time larger than 9 Gyrs)
characterized by a steep increase in mass, which corresponds to the initial
intense burst of star formation shown in the top right panel. The stellar mass
stays almost constant between 8 and 6 Gyrs ago, and then increases rapidly down
to present time. This behaviour is in quite good agreement with predictions
from the {\it SAM} run. In contrast, the {\it SAM2} run exhibits an
approximately linear accretion history, consistent with the approximately
constant star formation history shown in the top right panel. These accretion
histories are shown again in the lower right panel, together with the integrals
of the star formation histories shown in the top right panel. For the {\it SAM}
and {\it SAM2} runs, the total mass in progenitors matches the integrals of the
star formation histories (dotted and solid lines overlap perfectly). Indeed,
all the stars that are associated with the BCG at present in these runs, were
formed in BCG progenitors. This is not the case in the {\it SPH}
simulation, where the two curves start diverging below lookback time $\sim 10$
Gyrs. By $z = 0$, the integral of the star formation rate of all BCG
progenitors is responsible for only about half of the final stellar
mass. Interestingly, even if the star formation histories of BCGs predicted by
the simulation and the semi-analytic runs differ significantly, their integrals
agree within $\sim 20$ per cent for the {\it SPH} and {\it SAM2} run. For the
{\it SAM} run, the integral of the star formation history is about a factor two
larger than that measured from the {\it SPH} simulation and from the {\it SAM2}
run. At $z=0$, the integrated total mass of the BCG is $\simeq 1.7 \times
10^{13} \msun$, $\simeq 2.1 \times 10^{13} \msun$ and $\simeq 3.2 \times
10^{13} \msun$ for the {\it SPH}, {\it SAM}, and {\it SAM2}, respectively.

Part of the excess mass associated with the {\it SPH} BCG and not formed in its
progenitors, comes from {\it diffuse} star formation occurring in unresolved
galaxies which eventually merge with the BCG by redshift $z=0$ (see also
\citealt{Saro08B}). To compute the star formation rate associated with each
{\it SPH} galaxy, we have summed all the star particles in the galaxy under 
consideration with formation time lower than the time-interval between two 
adjacent snapshots. Galaxies that are formally unresolved and that form outside 
the FOF group of the main halo, will not appear in the galaxy merger tree. 
Therefore, their contribution is not taken into account when computing the star 
formation history shown in the top right panel of Figure \ref{fi:GAL0S2}. Their
stellar mass, however, contributes to the total masses shown in the bottom left 
panel, once these galaxies merge with the BCG. 

The largest contribution to the excess stellar mass not formed in the
{\it SPH} BCG progenitors, however, comes from stars that formed in
Type-1 satellites and that were tidally stripped and added to the
diffuse ICL associated with the main cluster. We note that {\small
  SUBFIND} does not separate the stars bound to the BCG from the
diffuse stellar component \citep[e.g.,][]{2007MNRAS.377....2M}, so
that the BCG `stellar mass' will include this component. To quantify
the contribution due to the diffuse stars, we have identified all
satellite galaxies of the {\it SPH} BCG and traced them back in time
until their stellar mass reaches a maximum. This time corresponds
approximately to the last time the galaxy was a central galaxy of its
own halo. We then sum up the differences between their maximum masses
and their masses at $z=0$, and obtain a total mass of $\sim 1.6 \times
10^{13} \msun$. Interestingly, this amount corresponds almost exactly
to the difference between the final stellar mass associated with the
{\it SPH} BCG at $z=0$ and the integral of its star formation
history. We caution, however, that {\it SPH} galaxies may be too
fragile \citep{2006MNRAS.373..397S}, boosting the significance of this
effect. {In particular, the larger fraction of stellar mass
  formed at early times in absence of any regulating feedback
  mechanism, is enhancing this effect, thus over-predicting the amount
  of ICL.}

%%%%%%%%%%%%%%%%%%%%%%%%%%%%%%%%%%%%%%%%%%%%%%%%%%%%%%%%%%%%%%%%%%%%%%%%%%%%%%%
\subsection{Evolution of satellite galaxies}

\begin{figure*}
  \centerline{ \hbox{ \psfig{file=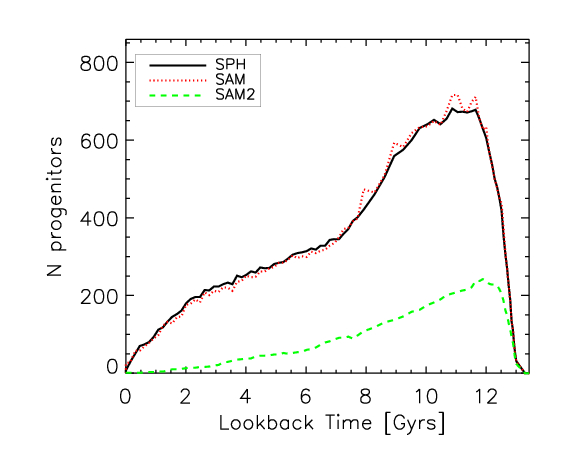,width=0.5\textwidth} \psfig{file=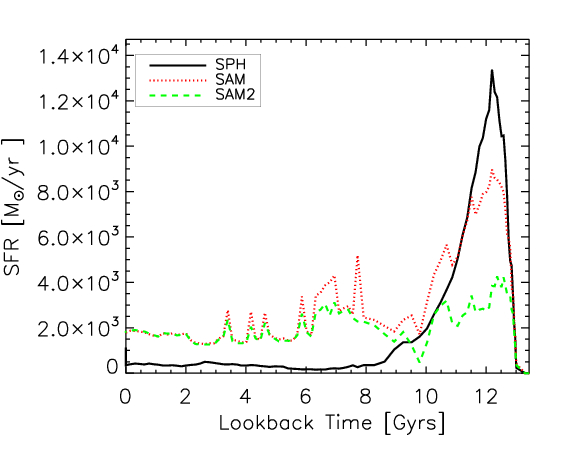,width=0.5\textwidth}
    }} \centerline{ \hbox{ \psfig{file=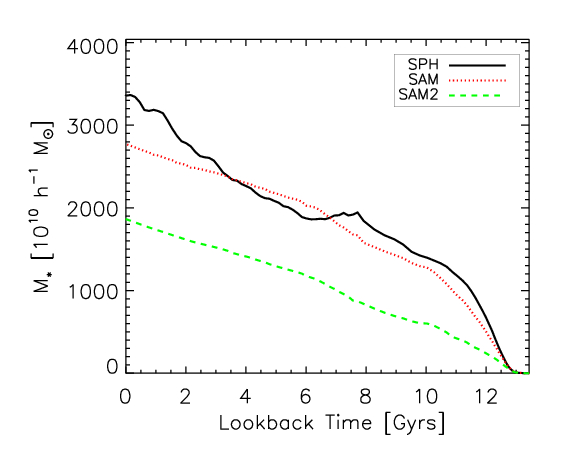,width=0.5\textwidth} \psfig{file=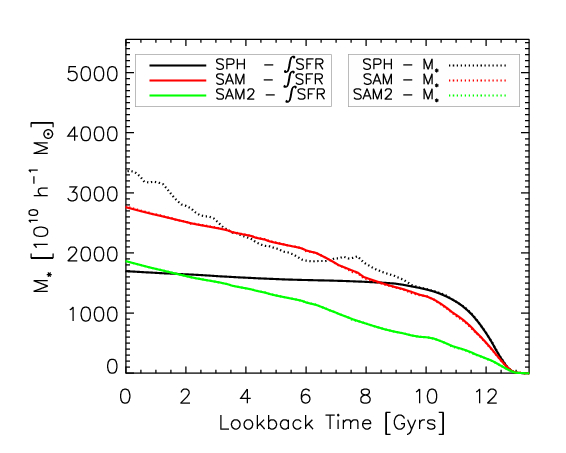,width=0.5\textwidth}
    }}
\caption{{\bf Top left panel}: number of BCG progenitors (galaxies that merge
  with the BCG by $z=0$) at different lookback times. {\bf Top right panel}:
  star formation history of the BCGs in each of the runs analysed in this study,
  obtained including the contribution from each BCG progenitor. {\bf Lower left
    panel}: total stellar mass in all BCG progenitors as a function of lookback
  time. {\bf Lower right panel}: same quantities shown in the lower left panel
  are compared with the integrals of the star formation histories shown in the
  top right panel. Dotted and solid lines for the {\it SAM} and {\it SAM2} runs
  overlap. Different colours correspond to different runs as indicated in each
  panel.}
\label{fi:GAL0S2}
\end{figure*}

\begin{figure}
\centerline{ \hbox{ 
\psfig{file=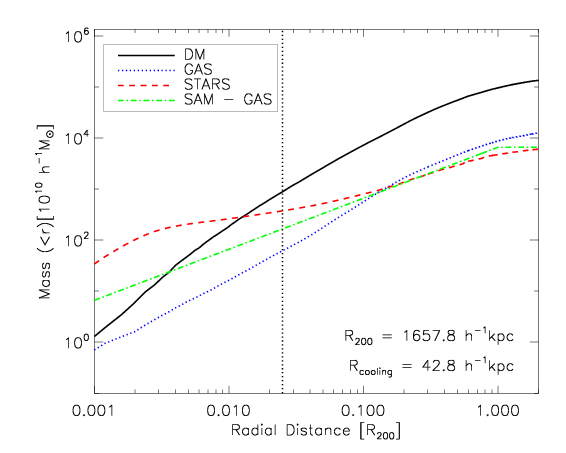,width=0.5\textwidth}
}} 
\caption{Total mass of dark matter (black solid line), gas (dotted
  blue line), and stars (dashed red line) enclosed within a given
  radius in the {\it SPH} simulation, at $z = 0$. The dot-dashed green
  line shows the gas profile assumed by the SAM adopted in this
  study. The vertical dotted line marks the cooling radius assumed in
  the SAM at this redshift (see text for details).}
\label{fi:prof_dens}
\end{figure}

\begin{figure*}
  \centerline{ \hbox{ \psfig{file=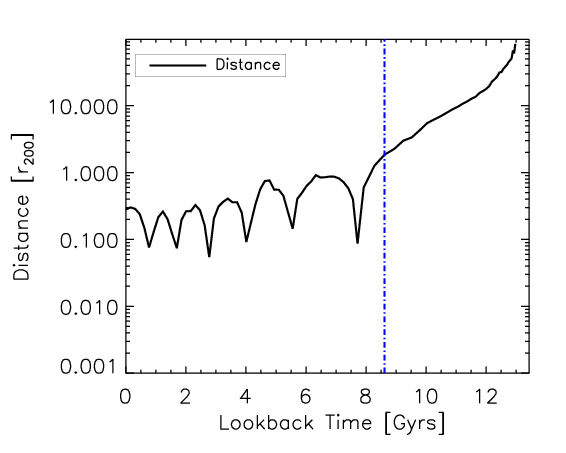,width=0.5\textwidth, height = 0.39\textwidth}
      \psfig{file=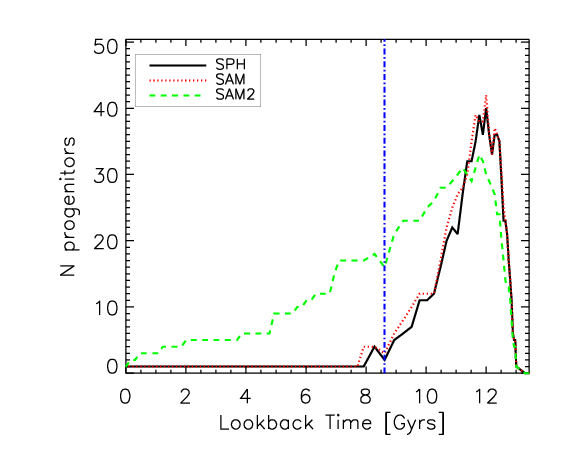,width=0.5\textwidth, height = 0.39\textwidth}
    }} 
  \centerline{ \hbox{ \psfig{file=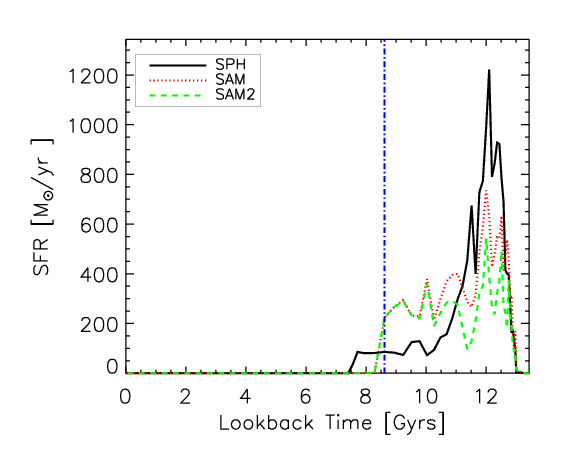,width=0.5\textwidth, height = 0.39\textwidth} \psfig{file=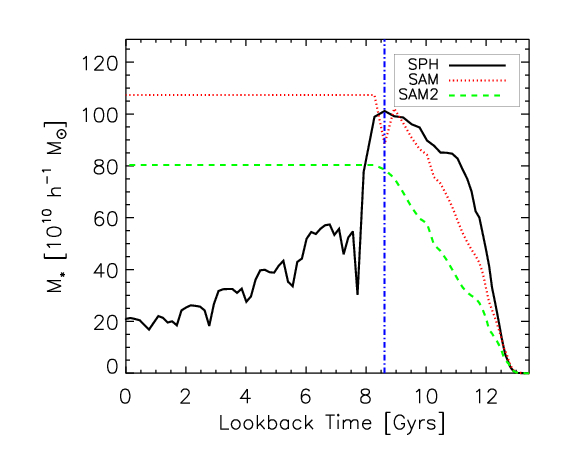,width=0.5\textwidth, height = 0.39\textwidth}
    }} 
\centerline{ \hbox{ \psfig{file=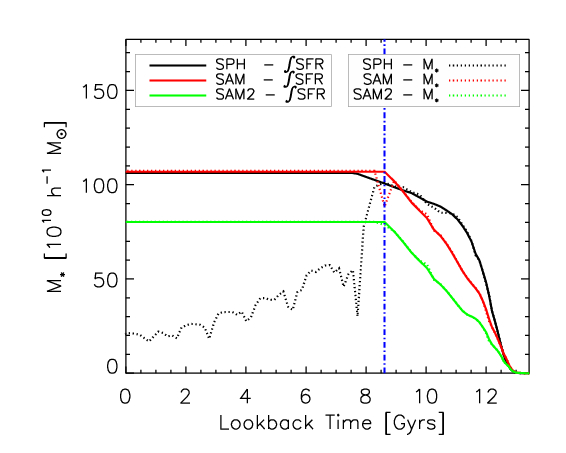,width=0.5\textwidth, height = 0.39\textwidth} \psfig{file=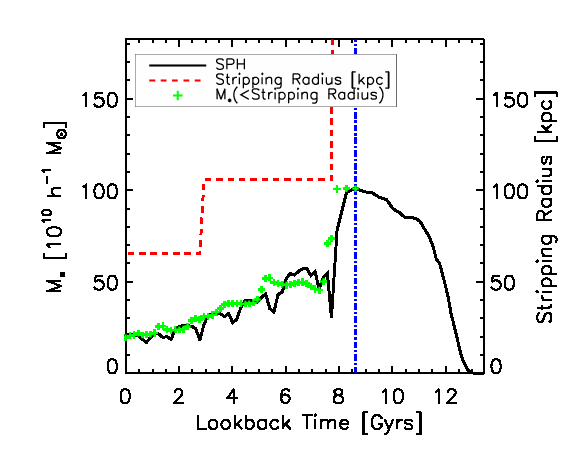,width=0.5\textwidth, height = 0.39\textwidth}
    }} 
\caption{Evolution of the sixth most massive satellite galaxy within $r_{200}$
  at $z = 0$, as predicted by the {\it SPH} simulation (solid black lines), by
  the {\it SAM} run (dotted red lines), and by the {\it SAM2} run (dashed blue
  lines). The merger trees of this galaxy are shown in
  Fig.~\ref{fi:TreeS2}. {\bf Upper left panel:} Distance between the main
  progenitor of the galaxy under consideration and the cluster centre as a
  function of lookback time, in units the cluster $r_{200}$ at the
  corresponding time. {\bf Upper right panel:} Number of progenitors as a
  function of lookback time. {\bf Middle left panel:} Total star formation
  history, obtained by summing up the contribution of all galaxy's
  progenitors. {\bf Middle right panel:} Total stellar mass in the galaxy's
  progenitors as a function of lookback time. {\bf Lower left panel:} Total
  stellar mass in the galaxy's progenitors and corresponding integrals of the
  star formation history.  {\bf Lower right panel}. Total stellar mass in all
  progenitors of the galaxy under consideration as measured from the {\it SPH}
  simulation (solid black line). The dashed line in this panel shows a
  simple estimate of the stripping radius (see text for details), and green
  crosses represent the total stellar mass of the {\it SPH} galaxy enclosed
  within this radius.  The vertical line in each panel marks the time when the
  galaxy becomes a satellite.}
\label{fi:GAL4S2}
\end{figure*}

Figure \ref{fi:GAL4S2} shows the evolution of the sixth most massive
satellite galaxy found in the {\it SPH} simulation at $z=0$, within
$r_{200}$. The merger trees of this same galaxy in the three runs
analysed in this paper were shown in Figure \ref{fi:TreeS2}. The upper
left panel shows how the distance between the main progenitor of the
galaxy and the cluster centre varies as a function of lookback
time. The vertical line in this (and in all the other panels) marks
the time when the galaxy becomes a satellite. This particular galaxy
has been orbiting around the cluster centre in the last $\sim
8.5$~Gyrs and is currently located at a distance of $\sim 0.3\,
r_{200}$ from the cluster centre. By $z=0$, the galaxy has completed
six orbits and both its apocentric distance and orbital period have
decreased as a consequence of dynamical friction. The upper right
panel of Figure \ref{fi:GAL4S2} shows the number of progenitors in the
three runs analysed in our study (as we did in Figure \ref{fi:GAL0S2},
for the BCG).  The figure shows that both in the {\it SPH} and in the
{\it SAM} run, there is only one merging episode after the galaxy has
become a satellite: it happens only $\sim 0.5$ Gyrs after the
satellite galaxy crossed $r_{200}$. In the {\it SAM2} run, the number
of progenitors decreases more gradually than in the other two runs, as
a consequence of the continuous accretion of Type-2 galaxies.

The mid-left panel of Figure \ref{fi:GAL4S2} shows the star formation history
of the galaxy under consideration.  All three runs are characterised by an
intense burst of star formation at early times, and no star formation in the
last $\sim 7$~Gyrs. As for the BCGs, however, the {\it SAM2} galaxy is
characterized by a less intense and broader initial burst, compared to its
{\it SPH} and {\it SAM} counterparts. Interestingly, the star formation
associated with the {\it SPH} simulated galaxy, lasts for $\sim 1$ Gyrs after
the galaxy has became a satellite. Since all available cold gas is
instantaneously turned into stars, new fresh material for star formation must
come from cooling on satellite galaxies occurring in the {\it SPH} simulation.
In the {\it SAM} and {\it SAM2} runs, no cooling is allowed on satellite
galaxies and all gas available is instantaneously turned into stars. So, by
construction, no star formation occurs after the galaxy is accreted onto a
larger structure.

The mid-right panel of Figure \ref{fi:GAL4S2} shows the total mass in all the
progenitors of the galaxy under consideration, as a function of lookback
time. The dip in the total stellar mass of the {\it SAM} galaxy at the position
of the vertical line is due to the adopted merger tree construction algorithm.
As explained in Section \ref{sec:Results1S2}, the algorithm looks for a
descendant of each halo in the two following snapshots so that in the dark
matter merger trees, haloes are allowed to skip a snapshot. This also happens
for the galaxy that is sitting at the centre of this structure, as in Figure
\ref{fi:GAL4S2}. In the lower left panel, the total mass in all progenitors is
compared to the integrals of the star formation histories, as we did for the
BCGs in the lower right panel of Figure \ref{fi:GAL0S2}. As for the BCGs, the
total mass for the {\it SAM} and {\it SAM2} galaxies overlap perfectly the
integral of the corresponding star formation history, confirming that all stars
in the final galaxy have been formed in its progenitors. The behaviour of the
{\it SPH} galaxy is more complicated, and we can identify two different
regimes:
\begin{itemize}
\item at early times, before the galaxy is accreted onto a larger structure,
  its stellar mass follows closely the integral of the star formation history.
\item as the galaxy crosses the virial radius and becomes a satellite, it
  starts loosing stellar mass because of tidal interactions with the cluster
  halo.
\end{itemize}
To better understand the mechanism which causes the measured loss of stellar
material from the {\it SPH} satellite galaxy, we define at each snapshot a
representative `stripping radius'. This quantity is defined, at each snapshot,
as the minimum cluster-centric distance reached by the orbiting satellite
galaxy by the time corresponding to the analysed snapshot. This simple estimate
of the stripping radius is shown by a red dashed line in the lower right panel
of Figure \ref{fi:GAL4S2}, together with the predicted enclosed mass, as a
function of lookback time.  The figure shows that the galaxy suffers a
significant loss of stellar mass during the first orbit, followed by smoother
but constant stripping down to $z=0$. Our estimate of the stripping radius is
clearly simplified, but it captures well the evolution of the bound stellar
mass as measured from the {\it SPH} simulation. At pericentres, {\small
  SUBFIND} tends to under-predict the bound stellar mass with respect to the
estimate obtained using the stripping radius defined above. This bias is
intrinsic in the {\small SUBFIND} algorithm, but cannot account for the total
amount of stellar mass loss at the apocentre, as discussed in
\citet{2007MNRAS.376..180N} and \citet{2009MNRAS.396.1329M} and produces an
artificial decrease in mass when a galaxy orbits close to the cluster
centre. This artificial loss of stellar material is {\it recovered} when the
galaxy travels to larger distances.

It is woth noticing that tidal stripping does not represent the main
contribution to the formation of the intra-cluster component, whose production
is mainly contributed by mergers associated with the formation of BCGs in
groups of galaxies \citep[e.g.,][]{2006ApJ...648..936R,2007MNRAS.377....2M}. As
noted earlier, {\small SUBFIND} is not able to separate the bound galaxy
stellar mass from the associated diffuse component. Therefore, at the time the
galaxy is accreted onto a larger structure, the galaxy's stellar mass is a
combination of its genuine stellar mass and of its own diffuse light. Tidal
interactions are more efficient in stripping the galaxy's diffuse component, as
this material resides at larger radii and is less bound. For the galaxy shown
in Figure \ref{fi:GAL4S2}, the minimum stripping radius ($\sim 65$ kpc) is
larger than the galaxy's radius, which suggests that the majority of the
stellar material lost by the {\it SPH} satellite galaxy while orbiting in the
cluster potential well is diffuse light of the group that contained the galaxy
before it was accreted onto the cluster.

We stress that our simplified estimate of the stripping radius is not
meant to provide a `model' to include in semi-analytic models of
galaxy formation. Indeed, tidal stripping of stars in SPH simulations
might be affected by numerical issues, and it has been demonstrated
that no numerical convergence has been achieved over the range of
resolutions examined \citep[][]{2007MNRAS.377....2M}

%%%%%%%%%%%%%%%%%%%%%%%%%%%%%%%%%%%%%%%%%%%%%%%%%%%%%%%%%%%%%%%%%%%%%%%%%%%%%%%
\subsection{Effect of the environment}

In this section, we study the effect of environment on predictions by
the {\it SPH} and the semi-analytic model used in our study, by
comparing a Type-0 central galaxy with a Type-1 satellite galaxy. In
order to carry-out this comparison, we choose two galaxies with
similar stellar mass at $z = 0$ of about $3 \times 10^{11} \msun$.

\begin{figure*}
  \centerline{
    \hbox{
      \psfig{file=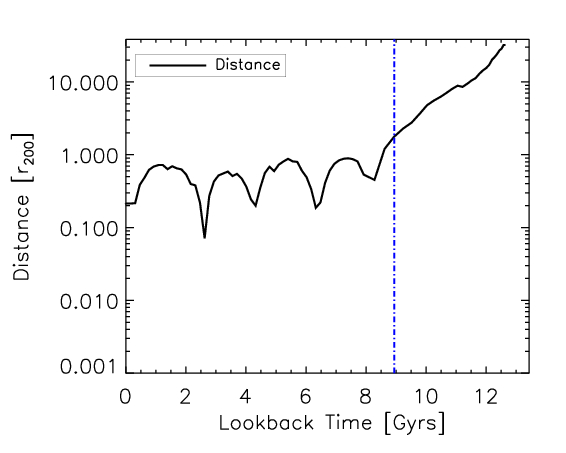,width=0.5\textwidth} 
      \psfig{file=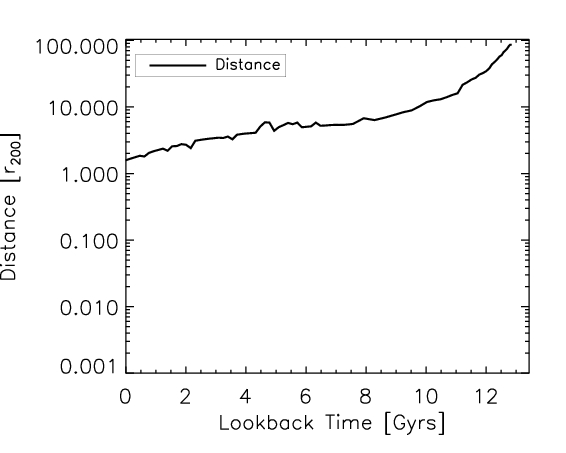,width=0.5\textwidth} 
      }}
  \centerline{
    \hbox{
      \psfig{file=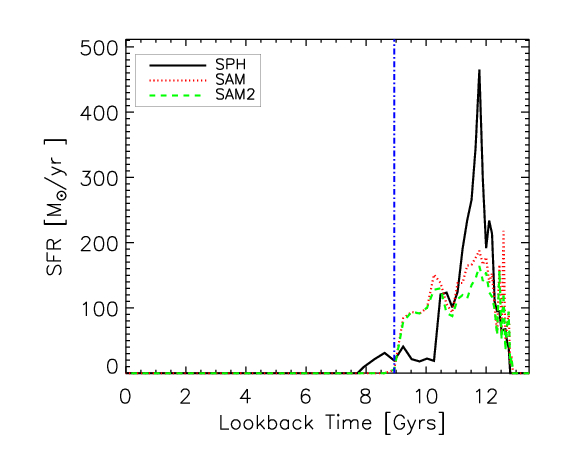,width=0.5\textwidth} 
      \psfig{file=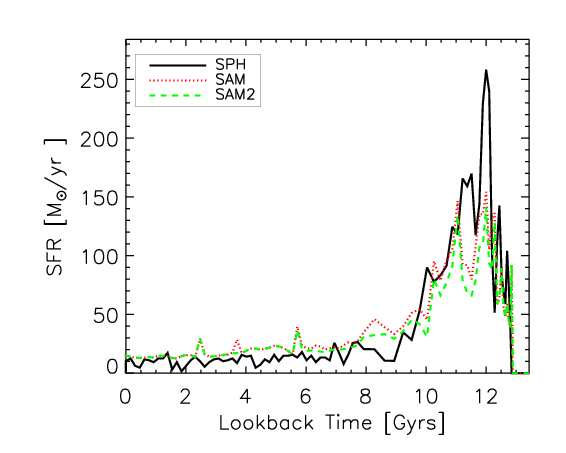,width=0.5\textwidth} 
      }}
  \centerline{
    \hbox{
      \psfig{file=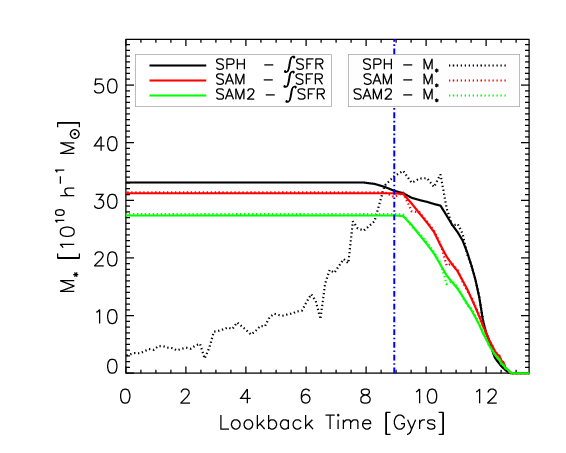,width=0.5\textwidth} 
      \psfig{file=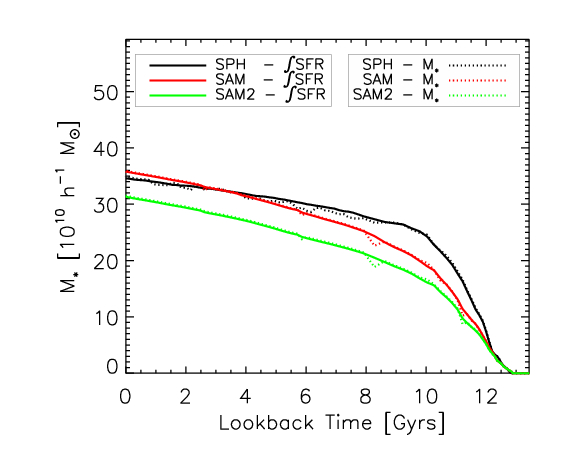,width=0.5\textwidth} 
      }}
\caption{Evolution of a central Type-0 galaxy (right panels) and of a satellite
  Type-1 galaxy (left panels) of similar mass ($\sim 3\div 3.5\times 10^{11}
  \msun$) at $z=0$. {\bf Upper panels:} cluster-centric distance of the main
  progenitor of the galaxies under consideration as a function of lookback
  time, in units of $r_{200}$ at the corresponding time. {\bf Central panels:}
  total star formation rate in all progenitors as a function of lookback
  time. {\bf Lower panels:} total stellar mass (dotted black line for the {\it
    SPH} run, dotted red line for the {\it SAM} run, and dotted green line for
  the {\it SAM2} run) in all progenitors, and integral of the corresponding
  star formation history (solid lines) as a function of lookback time. The
  dot-dashed vertical blue line in the left panels marks the lookback time
  corresponding to the transition from central to satellite. }
\label{fi:Same2S2}
\end{figure*}

The evolution of the central Type-0 galaxy is shown in the right panels
of Figure \ref{fi:Same2S2}, while the left panels show the
corresponding evolution of the satellite Type-1 galaxy.  As in Figure
\ref{fi:GAL4S2}, the vertical line in the left panels marks the lookback times
corresponding to the transition from central to satellite. The upper
panels of Figure \ref{fi:Same2S2} show the cluster-centric distance of
the galaxy's main progenitor as a function of lookback time. At $z=0$,
the central galaxy is located at a distance of $\sim 1.5 \, r_{200}$
from the cluster centre. The satellite galaxy crosses the cluster
virial radius at lookback time $\sim 9$ Gyrs, completes 5 orbits
around the cluster centre with an average period of $\sim 1.5$
Gyrs, and lies at $\sim 0.2 \,r_{200}$ from the cluster centre at
present.

The central panels of Figure \ref{fi:Same2S2} show the star formation histories
for the two galaxies under consideration in each of the three runs analysed in
this study. In all cases, the {\it SPH} galaxies exhibit a sharper and more
intense initial burst of star formation with respect to the {\it SAM} and {\it
  SAM2} corresponding galaxies, as found for the BCGs. The three runs are in
quite good agreement (much better than for the BCG shown in Figure
\ref{fi:GAL0S2}) for the Type-0 central galaxy analysed here. All three runs
predict a more intense and shorter episode of star formation for the satellite
galaxy than for the central galaxy. This is a consequence of our choice to
select galaxies with the same final stellar mass. For the central galaxy, the
level of star formation is low and approximately constant (around $10 \sfr$)
during the past 8 Gyrs, while no stars are formed in the satellite galaxy over
the same time-interval. As for the satellite galaxy shown in Figure
\ref{fi:GAL4S2}, the star formation in the {\it SPH} satellite lasts for about
one Gyr after the galaxy has been accreted onto the main halo. The closer
agreement between the {\it SAM} and {\it SAM2} predictions for the central
galaxy (right panel) is due to the fact that they lie in a `low density'
environment, in which the contribution from the Type-2 galaxy population is less
significant than in the main cluster. In addition, in these low-density
regions, tidal stripping plays a less important role, leading to a better
agreement between the {\it SPH} and the {\it SAM} results for the central
Type-0 galaxies.

The lower panels of Figure \ref{fi:Same2S2} show the total stellar
mass in all progenitors of the two galaxies and the integral of the
corresponding star formation history, as done in Figures
\ref{fi:GAL0S2} and \ref{fi:GAL4S2}. For the central Type-0 galaxy,
the mass increases most rapidly in the {\it SPH} simulation. The mass
growth is somewhat slower in the {\it SAM} run, but the predicted
final stellar mass is very close to the {\it SPH} prediction. In the
{\it SAM2} run, the stellar mass grows more slowly and the final
stellar mass reaches a lower value than those predicted by the {\it
  SAM} run and by the {\it SPH} simulation. For the Type-1 satellite
galaxy, the evolution is similar to that shown in the lower right
panel (for the central Type-0 galaxy) before the time of
accretion. After that time, the stellar mass of the {\it SAM} and {\it
  SAM2} galaxies remains almost constant, while the stellar mass of
the {\it SPH} satellite decreases continuously down to present time,
with a first important drop in stellar mass associated with the first
pericentric passage, and a smoother decrease of stellar mass during
the following orbits. By $z=0$, this particular galaxy has lost more
than $90$ per cent of its stellar mass at the time of accretion.

Summarising, we find a good agreement between the {\it SPH} and {\it SAM}
predictions for central galaxies lying in low-density environments, whereas a
different behaviour is found for the satellite galaxies. In particular, tidal
stripping of stellar material (a process which is not implemented in the
adopted semi-analytic model) is playing a crucial role in determining the final
mass of satellite galaxies.

%%%%%%%%%%%%%%%%%%%%%%%%%%%%%%%%%%%%%%%%%%%%%%%%%%%%%%%%%%%%%%%%%%%%%%%%%%%%%%%
\section{Discussion}
\label{sec:disc}

Our results demonstrate that predictions of a stripped-down version of the
semi-analytic model used in our study are in {\it quite nice agreement} with
results from {\it SPH} simulations including the same physics, {\it in
  `average' density environments. In higher density regions}, like galaxy
clusters, {\it predictions from simulations and semi-analytic models differ
  significantly, for physical reasons that are not specific of the model and
  the simulation used in this work}.  

At first sight, these considerations appear to be in conflict with conclusions
found in previous work (\citealt{Benson_et_al_2001},
\citealt{Yoshida_et_al_2002}, \citealt{2003MNRAS.338..913H},
\citealt{Cattaneo_et_al_2007}). All these studies have carried out comparisons
between simulations and stripped-down versions of different semi-analytic
models. In particular, Benson et al. (2001) compared the statistical properties
of galaxies found in the semi-analytic model developed by the Durham group
(they used the model presented by \citealt{ColeAl00}) with SPH simulations. The
semi-analytic model used in their study adopted a Monte Carlo technique for
building dark matter merger trees. Therefore, it was not possible to carry-out
a one-to-one comparison between model and simulation results. A comparison on
an object-by-object basis between the same semi-analytic model and simulations
was later carried out by Helly et al. (2003). Yoshida et al. (2002) compared
predictions from a SPH simulation of a galaxy cluster with those from the
Munich semi-analytic model (the implementation described in
\citealt{Kauffmann_etal_1999}). More recently, Cattaneo et al. (2007) compared
the galaxy population from a SPH simulation with predictions from a
stripped-down version of the GalICS model \citep{2003MNRAS.343...75H}. All
these studies agree that the two techniques provide results that are
statistically consistent.

It is important to stress that our results are not in contradiction with the
above mentioned studies. When focusing on a statistical comparison between
model and simulation predictions, we also get results that are broadly
consistent (see e.g. Figures \ref{fi:Mass_comparison_radiusS2} and
\ref{fi:MFSS2}). Although this is the message that has been generally accepted
by the community, we stress that the tension between models and simulations
discussed in our paper is also present in previous studies when focusing on an
object-by-object comparison. Yoshida et al. (2002) already pointed out that
when merging times are computed using the classical dynamical friction formula,
semi-analytic models tend to over-predict the number of galaxies with respect
to hydrodynamical simulations. A similar conclusion can be inferred by Fig.~4
in Cattaneo et al. (2007). Fig.~8 of the same paper shows that the
semi-analytic model predicts larger star formation rates than SPH simulations
at low redshift, as we have shown above. Yoshida et al (2002) also noted that
some gas is continuing to cool onto the galaxies after they are accreted in a
larger halo, in the hydrodynamical simulation.  Finally, we note that Benson et
al. (2001) already pointed out that the main reason for the differences they
found between their stripped-down semi-analytic model and the SPH simulation
was that gas cools more efficiently in massive haloes at early times in the SPH
simulation, as we have discussed in Section \ref{sec:bcgs}.

Hydrodynamical simulations of galaxy formation are becoming increasingly
sophisticated. The continuous increase of computational power will also allow us
to run simulations with higher and higher resolution. It will, however, remain
the necessity to use `sub-grid' physics, at least until we will not be able to
build a complete theory for the various physical processes that drive galaxy
formation and evolution. Semi-analytic models can access a large dynamic range
of mass and spatial resolution and allow a fast exploration of the parameter
space and of the influence of different physical assumptions. They will
therefore remain a valid method to study galaxy formation for the foreseeable
future. Given the complementarity between the two techniques, it is important
to analyse in more detail any disagreement between model and simulation
predictions, so as to test the robustness of numerical predictions and possibly
improve semi-analytic calculations. 

Quoting Benson et al. (2001), {\it the main conclusion of this paper is that
  the agreement between the SPH simulation and the stripped-down version of the
  semi-analytic model is better than a pessimist might have expected.} A
detailed comparison on an object-by-object basis, however, clarifies that there
are important discrepancies between predictions from the two techniques,
identifying areas where further work is necessary in order to improve our
galaxy formation models.

%%%%%%%%%%%%%%%%%%%%%%%%%%%%%%%%%%%%%%%%%%%%%%%%%%%%%%%%%%%%%%%%%%%%%%%%%%%%%%%
\section{Summary} 
\label{concS2}

In this paper, we have carried out a comparison between the cluster
galaxy population predicted by hydrodynamical SPH simulations and that
predicted by a stripped-down version of a semi-analytic model. Our
semi-analytical model and our simulation consider only cooling and a
simple prescriptions of star formation, which consists in transforming
instantaneously any cold gas available into stars. In addition, we
have considered both the case when satellite galaxies merge
instantaneously after the mass of the parent subhalo falls below the
resolution limit of the simulation, and the case when these galaxies
are assigned a residual dynamical friction merging time. By
construction, the former implementation can be compared directly with
merger trees extracted from the hydrodynamical simulation used in our
study. We stress that our results are not meant to be compared with
observational data. Rather, the aim of our study is to carry out a
detailed comparison between two techniques that are widely used to
study galaxy formation and evolution in a cosmological framework, at
the stripped-down level considered in this study. It is important to
note that, given the limitations of these techniques, neither of them
is likely providing the `correct' answer. By exploring the reasons for
disagreements, however, we are able to identify areas where further
work is needed. The main results of our study can be summarised as
follows:

\begin{enumerate}
  
\item For central galaxies, the agreement between predictions from the
  simulation and the semi-analytic model  is quite good outside the
  cluster environment. For the BCG, the final stellar mass and mass accretion
  history predicted by the two techniques are also comparable. The predicted
  star formation histories, however, differ significantly. In particular, the
  {\it SPH} BCG exhibits a lower level of star formation activity at low
  redshift, and a more intense and shorter initial burst of star formation with
  respect to the {\it SAM} prediction.

\item The higher level of star formation activity predicted by the
  semi-analytic model at low redshift, is due to the assumption of an
  isothermal gas distribution, which leads to much larger cooling
  rates in the semi-analytic model at late times. When neglecting
  feedback, as we are doing in this study, gas particles in SPH
  simulations distribute according to a profile that is not well
  described by an isothermal distribution because cooling at high
  redshift efficiently removes gas from the inner regions of a galaxy
  cluster, and leaves small amounts of gas available for cooling at
  lower redshift. We stress that it is not useful to `tune' the SAM
  model to reproduce the SPH results because in a more realistic
  simulation including feedback, the gas profile would be again
  modified.

\item Only about half of the final stellar mass of the {\it SPH} BCG was formed
  in its progenitors. The other half is contributed material associated with
  galaxies that were accreted onto the cluster halo. These star particles will
  later represent the largest fraction of the diffuse stellar component
  associated with the {\it SPH} BCG itself.

\item {\it SPH} satellites can loose large fractions of their stellar mass (up
  to 90 per cent of the stellar mass at the time of accretion), due to tidal
  stripping. This process is not included in the semi-analytic model adopted in
  this study, leading to satellite masses which are systematically larger than
  the corresponding values found in the simulation.

\item In the simulation, some cooling occurs on satellite
  galaxies. This can last for up to 1 Gyr after accretion but is,
  however, important only for the most massive satellites. Gas cooling
  on satellite galaxies is not included in the model used in our
  study, and in most of the semi-analytic models discussed in the
  recent literature (for a first attempt to include this process, see
  \citealt{2008MNRAS.389.1619F}). More work is, however, needed to
  clarify how this would be affected by the inclusion of a regulating
  feedback process in hydrodynamical simulations.

\end{enumerate}

As discussed above, the discrepancies found between semi-analytic predictions
and simulation results, identify specific areas where further work is needed in
order to improve our galaxy formation models. This will ultimately help us to
construct better tools that can assist us in understanding the physical
processes driving galaxy formation and evolution.

%%%%%%%%%%%%%%%%%%%%%%%%%%%%%%%%%%%%%%%%%%%%%%%%%%%%%%%%%%%%%%%%%%%%%%%%%%%%%%%
\section*{Acknowledgements}
We thank Volker Springel for making available the substructure finder and
merger tree construction software that was originally developed for the
Millennium Simulation project. We thank J\'er\'emy Blaizot and Pierluigi Monaco
for useful discussions. AS acknowledges the receipt of a Marie Curie Host
Fellowships from the EARA-EST programme, and the hospitality of the
Max-Planck-Institut f\"ur Astrophysik where this project was initiated. GDL
acknowledges financial support from the European Research Council under the
European Community's Seventh Framework Programme (FP7/2007-2013)/ERC grant
agreement n. 202781. This work has been partially supported by the INFN-PD51
grant, by a PRIN-INAF Grant, by the ASI-COFIS Grant and by the PRIN-MIUR Grant
``The Cosmic Cycle of Baryons''. Numerical computations have been performed
performed at CINECA (``Centro Interuniversitario del Nord Est per il Calcolo
Elettronico''), with CPU time assigned thanks to an INAF-CINECA grant.

\bibliographystyle{mn2e}
\bibliography{references}

\end{document}